\documentclass[12pt]{iopart}
\usepackage{iopams,graphicx}
\usepackage{lscape}

\begin{document}

\title{Lookback time and Chandra constrains on cosmological parameters}

\author{D. Tonoiu, A.Caramete, L.A. Popa} {
\address{Institute for Space Sciences Bucharest-Magurele, Ro-077125 Romania}}
\eads{\mailto{tonoiud@spacescience.ro},
\mailto{acaramete@spacescience.ro}, \mailto{lpopa@spacescience.ro}}

\begin{abstract}

 In this paper we combine the WMAP7 with lookback time and Chandra gas fraction data to constrain the main cosmological parameters and the equation of state for the dark energy. We find that the lookback time is a good measurement that can improve the determination of the equation of state for the dark energy with regard to other external data sets. We conclude that larger lookback time data set will further improve our determination of the cosmological parameters.

\end{abstract}


\section{Introduction}

 The seventh-year Wilkinson Microwave Anisotropy Probe (WMAP) data rigorously test the standard cosmological model placing constraints on its basic parameters. The WMAP measurements alone are not enough to break the degeneracy among some cosmological parameters or to place constraints on non-standard cosmological models. For example, measurements of cosmic microwave background (CMB) power spectrum alone do not strongly constrain the curvature of the universe characterized by the energy density parameter $\Omega_k$. One needs to complement the CMB data with the luminosity or angular diameter distances measurements in order to constrain $\Omega_k$ because the astrophysical distances depend also on the expansion history of the universe \cite{komatsu-wmap7}.

 There are conclusive evidences that the universe is in a state of accelerated expansion. The Hubble diagram of Type Ia supernovae (SNeIa, \cite{Hicken09, schaf09, guimaraes09}), combined with  CMB anisotropy measurements (\cite{komatsu-wmap7, dunkley09}), baryon acoustic oscillations (BAO) from the galaxy distribution data (\cite{percivaletal07, samushia09a, gaztanaga09, wang09}), and galaxy cluster gas mass fraction measurements (\cite{allen08, samushia08, ettori09}) support the idea that we live in a spatially-flat universe where nonrelativistic matter make almost 30\% of the critical energy density while the rest is an unknown component called dark energy, with negative effective pressure being responsible for the present phase of accelerated  expansion of the universe \cite{samushia09}.

 Constrains on dark energy density parameter $\Omega_\Lambda$ and on its equation of state $w$ (the ratio of pressure to energy density), can explain the nature of the repulsive force causing the acceleration of the universe. One possible explanation for this unknown component is an energy density constant in time and uniform in space. Such a cosmological constant ($\Lambda$) was originally postulated by Einstein to explain a static universe, later rejected when the expansion of the Universe was first detected and presently reinstated to account for the dark energy. Still, the computed value of $\Lambda$ is expected to be $10^{120}$ larger than the observed one. Another cosmological scenario considers that dark energy is a dynamical scalar field with a time varying equation of state. An alternative explanation of the accelerating expansion of the Universe is that general relativity or the standard cosmological model is incorrect.

 However, ground and space observations can not discriminate among different dark energy scenarios as the correct explanation of the observed accelerating universe: a cosmological constant, a dynamical scalar field or a modification of general relativity  \cite{DETF, peebles02}.

 The WMAP7-year data combined with other astrophysical measurements \cite{komatsu-wmap7} place constraints on the dark energy. Assuming a flat universe ($\Omega_k \sim 0$), an accurate determination of the Hubble expansion rate ($H_0$) helps in improving the limit of the equation of state of the dark energy \cite{spergel03, hu05}. In the paper of Komatsu et. all. \cite{komatsu-wmap7} from the joint analysis of WMAP7+BAO+$H_0$ in the case of a time independent equation of state, a value of $w = -1.10 \pm 0.14$ at $68\%$ CL was obtained. Furthermore, adding high-z supernova data to their analysis a more stringent limit was obtained, $w = -0.98 \pm 0.053$ at $68\%$ CL. However, this later result does not take into account the systematic errors in supernovae, which are comparable with the statistical errors \cite{kessler09, hicken09}. Also, combining the cluster abundance and 5-year WMAP data, Vikhlinin et al. \cite{vikhlinin09} found that for a flat universe $w = -1.08 \pm 0.18$ at $68\%$ CL. Furthermore, adding BAO \cite{eisenstein05} and the supernova data \cite{davis07}, they found $w = -0.991 \pm 0.09$ at $68\%$ CL.

 In this paper, we perform a joint analysis of the CMB-WMAP7 data, constraints on Hubble expansion rate inferred from the age of astrophysical objects using lookback time method (LBT) and measurements of the gas mass fraction of relaxed clusters from Chandra X-ray observatory(Chandra). Our aim is to investigate for several parameters of a given cosmological model which combination of this three data sets puts better constrains.

 We choose the lookback time method because it has the advantage of using the ages of distant objects which are independent of each other so we can avoid biases present in techniques that use distances of primary or secondary indicators in the cosmic distance ladder method \cite{samushia09}.

 Moreover, we use Chandra measurements because they currently provides one of the best constraints on $\Omega_m$ and have the advantage of being remarkably simple and robust in terms of its underlying assumptions \cite{allen08}.

 The paper is structured as follows: in section 2 we briefly present the lookback time method, in section 3 Chandra gas fraction experiment, in section 4 we describe the statistical analysis and present the data sets and finally, in the last sections, the results and conclusions.

%%%%%%%%%%%%%%%%%%%%%%%%%%%%%%%%%%%%%%%%%%%%%%%%%%%%%%%%%%%%%%%%%%%%%%%%%%%%%%%%%%%%%%%%
\section{Lookback time method}

  This time-based method uses ages of astrophysical objects (passively evolving galaxies or clusters of galaxies) which are independent of each other, to constrain cosmological parameters. The advantage of this method is that it avoids biases that are present in cosmic distance ladder method, offering an independent way to cross-check cosmological constraints obtained by using other methods \cite{samushia09a}.

 The lookback time is defined as the difference between the present age of the Universe ($t_0$) and its age at redshift $z$, $t(z)$,
\begin{equation}
t_{L}(z) = t_{0} - t(z) = {\int_0^{z}{{dz^{'}} \over {(1+z^{'})H(z)}}}
\end{equation}

where $H(z)$ is the Hubble expansion rate at redshift $z$.

 We use the spatially-flat cosmological standard model XCDM, X denoting the fact that the dark energy component has an time independent unknown equation of state. In this model, the Hubble expansion rate as a function of redshift can be written as:
\begin{equation}
H(z)\, =H_0 \,[ \Omega_{\rm m}(1 +z)^3 + (1 - \Omega_{\rm m}  )(1 +z)^{3(1+w)} ]^{1/2},
\end{equation}

\noindent where $\Omega_{\rm m}$ is the matter energy density parameter and $w$ is the equation of state for the dark energy.
%.........................................
\begin{figure}[!hbtp]
\centering
\includegraphics[angle=0, width=0.8 \textwidth]{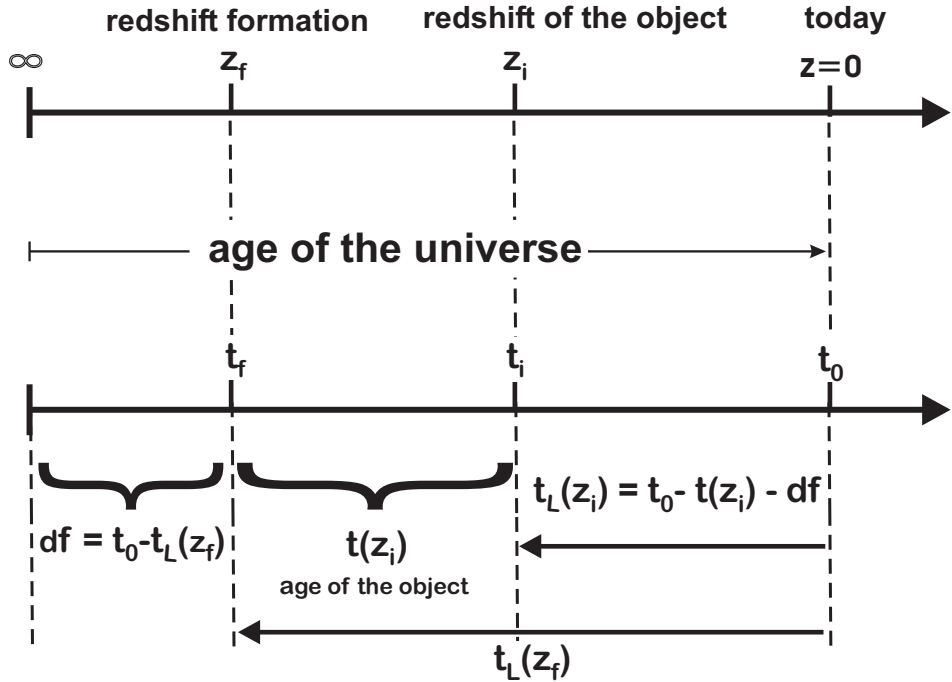}
\caption[Short]{\small The time evolution of the lookback time and the age of the universe as a function of redshift \cite{pires06} (see also the text).}
\end{figure}

 The observed lookback time $t_L^{obs}(z_i)$ (see the diagram presented in Fig. 1), to an object $i$ at redshift $z_i$ is defined as:
\begin{equation}
t_{L}^{\rm obs}(z_i)= t_0^{\rm obs}- t_i(z_i)-df.
\end{equation}
\noindent where $t_0^{\rm obs}$ is the measured age of the universe and $t_i(z_i)$ is the age of the object (a passively evolving galaxy or a cluster) defined as the difference between the age of the Universe at redshift $z_i$ and the age of the universe at the redshift of object formation, $z_f$:
\begin{equation}
t_i(z_i)=t(z_i) - t(z_f) =  t_{L}(z_f) \, - \, t_{L}(z_i)
 =  \int_{z_{i}}^{z_f}{dz'\over(1 + z')H(z')},
\end{equation}

 The third term in equation (3), $df$ is a delay factor that encodes our ignorance regarding the formation redshift of the object $z_f$ and is defined as: $df = t_0^{\rm obs}-t_L({z_F})$.

 The lookback time method has the important feature that the age of distant objects are independent of each other and we can avoid biases present in techniques that use distances of primary or secondary indicators in the cosmic distance ladder method \cite{samushia09a}.
%..........................................

\section{Chandra X-ray cluster gas fraction}

It is shown that the gas fraction(fgas) in X-ray luminous, dynamically relaxed clusters measured with the Chandra X-ray Observatory
, helps in better constraining the cosmological parameters\cite{allen08, allen04}. Following  \cite{white93, eke98}, the matter content of the largest clusters of galaxies provides an almost fair sample of the total matter of the universe. Measurements of the ratio of baryonic and total mass in clusters of galaxies should closely match the ratio of $\Omega_{\rm b}$ and $\Omega_{\rm m}$. The matter energy density parameter $\Omega_{\rm m}$ can be determined by combining measurements of the baryonic mass fraction in the largest galaxy clusters with accurate determinations of  $\Omega_{\rm b}h^2$ from cosmic nucleosynthesis and CMB experiments, and a reliable measurement of the Hubble constant, $H_0$.

  The X-ray clusters gas fraction model fitted to a reference $\Lambda$CDM cosmology is \cite{allen08}
\begin{equation}
f_{\rm gas}^{\rm \Lambda CDM}(z) = \frac{ K A \gamma b(z)} {1+s(z) }
\left( \frac{\Omega_{\rm b}}{\Omega_{\rm m}} \right)
\left[ \frac{d_{\rm A}^{\rm \Lambda CDM}(z)}{d_{\rm A}(z)} \right]^{1.5},
\end{equation}
\noindent where $d_{\rm A}(z)$ and $d_{\rm A}^{\rm \Lambda CDM}(z)$ are the angular diameter distances to the clusters in the current test model XCDM and the reference standard cosmological model $\Lambda CDM$.
\begin{equation}
d_A = { c \over H_0 (1+z)\sqrt{\Omega_k}} \sinh \left( \sqrt{\Omega_k} \int_0^z {H_0 dz \over H(z)} \right)
\end{equation}
 The systematic uncertainties in the Chandra experiment are parameterized by the angular correction A, the non-thermal pressure support in the clusters $\gamma$, the baryonic mass fraction in stars $s(z) = s_{0}(1+\alpha_{s}z)$, the bias factor $b(z) = b_0(1+\alpha_{b}z)$ and an overall calibration parameter K for the residual uncertainty in the accuracy of the instrument calibration and X-ray modeling.

\section{Analysis}

    We perform our analysis in the framework of the extended XCDM cosmological model described by 7 + 1 free parameters:
\begin{equation}
\Theta = (\underbrace{\Omega_b h^2, \Omega_{DM} h^2, \theta_s, \tau, w, n_s, A_s}_{standard}, df)
\end{equation}
assuming uniform priors for all the parameters.

 Here $\Omega_b h^2$ and $\Omega_{DM} h^2$ are the baryonic and dark matter energy density parameters, $\theta_s$ is the ratio of the sound horizon  distance to the angular diameter distance, $\tau$ is the reionization optical depth, $w = {p \over \rho}$ is the equation of state parameter ($p$ and $\rho$ are the pressure and energy density of the dark energy), $n_s$ is the scalar spectral index of the primordial density perturbation power spectrum, $A_s$ is its amplitude at the pivot scale and $df$ is the lookback time delay factor.
 Table 1 presents the parameters of our model, their fiducial values and the prior ranges adopted in the analysis.\\*
\begin{table}[!ht]

\caption{\small The parameters of our model, their fiducial values and the prior ranges adopted in the analysis}
\centering
\begin{tabular}{c c c c}
\\
\hline \hline
&Parameter &Fiducial value &Prior range \\ [0.5ex] % extra vertical spacing
\hline \\ [0.001ex]
$ $ & $\Omega_b h^2$ & $0.0223$  & $0.005 \rightarrow 0.1$\\
$ $ & $\Omega_{DM} h^2$ & $0.105$  & $0.01 \rightarrow 0.99$\\
$ $ & $\theta_s$ & $1.04$  & $0.5 \rightarrow 10$\\
$ $ & $\tau$ & $0.09 $ & $0.01 \rightarrow 0.8$ \\
$ $ & $w$ & $-1$  & $-2 \rightarrow 0$ \\
$ $ & $n_s$ & $0.95$  & $0.5 \rightarrow 1.5$ \\
$ $ & $ln[10^{10}A_s]$ & $3$  & $2.7 \rightarrow 4$ \\
$ $ & $df(Gyr)$ & $1.5$  & $0 \rightarrow 3$\\[1ex]
\hline
\end{tabular}
\label{table:params}
\end{table}

 We modified the CosmoMC Monte Carlo Markov Chain (MCMC) public package \cite{lewis02} for our extended $\theta$ parameter space to sample from the posterior distribution giving the following experimental datasets:

The \textbf{WMAP7} temperature and polarization CMB latest measurements \cite{komatsu-wmap7}

The \textbf{Lookback time (LBT)} measurements. From Ref. \cite{capozziello04} we use the ages of six galaxy clusters in the redshift range $0.10\le z\le 1.27$. The standard deviation uncertainty for this age measurements is about 1 Gyr. We also use the age of 32 passively evolving galaxies from Ref. \cite{simon05} in the redshift interval $0.117 \le \,z \, \le 1.845$. The error for this sample ($1 \sigma$) is $12\%$ of the age measurements. Therefore, we have 38 measurements of $t_i(z_i)$ with uncorrelated uncertainties $\sigma_i$. For the age of the Universe we use the WMAP estimate, $t_0^{\rm obs}=(13.69\pm 0.13)\ \rm Gyr$ \cite{dunkley09}.

 Then, we compute the $\chi^2$ function for each cosmological model described by the set of parameters given in equation (7) as:
\begin{equation}
\chi_{LBT}^2(\theta) = \sum_{i = 1}^{38}\frac{ (t_{\mathrm{L}}(\theta) - t_{\mathrm{L}}^{\rm obs}(z_i,df))^2}{ \sigma_{i}^2 +
\sigma_{t_0^{\rm obs}}^2} +\frac{ (t_{{0}}(\theta) - t_{{0}}^{\rm obs})^2}{\sigma_{t_0^{\rm obs}}^2},
\end{equation}

where $t_L$ and $t_0$ are the theoretical predicted values of the lookback time  and of the age of the universe and $t_L^{obs}$ and $t_0^{obs}$ are the corresponding measured values. Also, $\sigma_i$ is refereing to the one standard error of the experimental data and $\sigma_{t_0}^{\rm obs}$ is the uncertainty in the estimate of $t_0$. \\*

 \textbf{Chandra X-ray gas fraction}. We used fgas data following the work from Ref. \cite{allen08}, \cite{allen04} and \cite{rapetti05}. We modified the CosmoMC Monte Carlo Markov Chain public package \cite{lewis02} to allow the use of fgas data  when exploring the cosmological parameter space, implementing fgas module presented in Ref. \cite{RapWP}.

 Our implementation of this new module into CosmoMC package considers all seven parameters related to systematic uncertainties as being constant and having the following values: $A = 0.2$, $\gamma = 1.05$, $s_0 = 0.16$, $\alpha_s = 0$, $b(z) = 0.824$, $\alpha_b = 0$ and $K = 1$. This is different from the approach described in \cite{allen08}, where either Gaussian or linear uncertainties are taken into consideration and after that, the seven parameters are added to the original list of 13 possible free parameters. In this manner, we manage to decrease the computational costs, especially when trying to constrain non-standard models using also the Chandra data.

 The $\chi^2$ function we used for fgas data has the form:
\begin{equation}
\chi_{CH}^2(z) = \sum_{i = 1}^{42} \frac{(f_{gas}^{\Lambda CDM}(z) - f_{gas}^{exp}(z) )^2}{\sigma_{fgas}^{exp}(z)^2} + \frac{(\eta - 0.214)^2}{(0.022)^2}
\end{equation}
\noindent where $f_{gas}^{\Lambda CDM}$ is the gas fraction fitted to a reference $\Lambda CDM$ cosmological model and is given by (5) and $f_{gas}^{exp}(z)$, $\sigma_{fgas}^{exp}(z)$ are the experimental values of gas fraction and the associated errors. In the above equation, $\eta$ is the slope of the $f_{gas}(z)$ in a region with a radius for which the mean enclosed mass density is 2500 times the critical density of the universe at the redshift of the cluster, as measured for the reference $\Lambda CDM$ standard cosmological model \cite{allen08}. The difference between the experimental slope $\eta$ and its reference value is normalized to its expected standard deviation squared.
\\*
 WMAP7, LBT and Chandra data sets (WMAP7+LBT+Chandra) are combined by multiplying the likelihoods.
 We have performed a likelihood analysis using three cosmological data sets: CMB-WMAP7, lookback time and the X-ray cluster gas fraction.

\section{Results}
%..............................................................................

 In order to see, for several parameters of the chosen cosmological model XCDM, which data provides better constrains we run the modified CosmoMC package on a parallel computing system by using 64 independent chains for the following combinations of data sets: WMAP7, WMAP7+LBT, WMAP7+Chandra and WMAP7+LBT+Chandra. We impose for each case the Gelman \& Rubin convergence criterion \cite{gelman92}.

 We present in Table 2, the mean value and $1 \sigma$ error at a 68\% CL, obtained from our analysis, for the following cosmological parameters: $\Omega_b h^2$, $\Omega_{DM} h^2$, $\Omega_{\Lambda}$, $\Omega_{m}$, $n_{s}$, $w$, $H_0$, $age(Gyr)$, $log[10^{10} A_s]$. In the first column, we present the results for the reference WMAP-WCDM model\footnote{http://lambda.gsfc.nasa.gov/product/map/dr4/params/wcdm\_sz\_lens\_wmap7.cfm} and in the second column, our simulation with WMAP7 data alone. Then, in the next two columns we present the results obtain from the combination of WMAP7 with LBT and Chandra data and finally, in the last column, we show the results when all three data sets are joined. Comparing with the WMAP-WCDM reference model, we conclude that our computation best constrain the $\Omega_{DM} h^2$, $\Omega_{\Lambda}$, $\Omega_{m}$ parameters as expected, if we use all three data sets (WMAP7, LBT, Chandra) and the parameters $\Omega_b h^2$, $n_{s}$, $w$, $H_0$, $Age(Gyr)$, $log[10^{10} A_s]$ are better constrained with WMAP7 and Chandra data.

 In Fig. 2 we show the correlation between some cosmological parameters, ($\Omega_{m}, \Omega_{\Lambda}$), ($\Omega_m , w$), ($H_0, \Omega_m$), ($H_0, w$), ($Age/Gyr, w$), ($\Omega_{\Lambda}, w$). This are the joint two-dimensional marginalized distributions with 68\%, 95\%, 99\% CL for the following combinations of data sets: WMAP7 data only, WMAP7+LBT, WMAP7+Chandra and WMAP7+LBT+Chandra. The plane $\Omega_{m} - \Omega_{\Lambda}$ is better constrain when we combine all three data sets. Also for others correlations, the WMAP7 and Chandra combination is the optimal one. Although adding LBT data set to the analysis we obtain better constraints upon all cosmological parameters, the central values of the distributions are shifted when comparing with WMAP-WCDM reference model, so we prefer to use only the WMAP7+Chandra data sets when we want to see the correlation between the equation of state of dark energy and other cosmological parameters: $\Omega_m$, $H_0$, $age/Gyr$, $\Omega_{\Lambda}$ and also, the correlation between $H_0$ and $\Omega_m$.

 We have shown that, for some parameters of a cosmological model with an unknown time independent equation of state, adding LBT to Chandra and WMAP7 data, leads to an improvement of their error bars. In the future, we expect that a larger LBT data set will further improve the constrains on cosmological parameters.
%..............................................................................
\begin{landscape}
\begin{table}[!ht]
\caption{\small The mean and 1$\sigma$ uncertainty for cosmological parameters in WMAP7, LBT and Chandra combined data analysis}
%\centering
\begin{tabular}{c c c c c c}
\\ \\
\hline \hline
 & WCDM-WMAP7 & W7 & W7+LBT & W7+CHANDRA & W7+LBT+CHANDRA \\ [0.5ex] % extra vertical spacing
\hline \\ [0.001ex]
$\Omega_b h^2$ & ${0.02258}^{+0.00063}_{-0.00062}$  & ${0.02259}^{+0.00058}_{-0.00058}$ & ${0.02137}^{+0.00073}_{-0.00068}$ & $\mathbf{{0.02202}^{+0.00047}_{-0.00047}}$ & ${0.01909}^{+0.00046}_{-0.00046}$ \\
$\Omega_{DM} h^2$ & ${0.1112}^{+0.0058}_{-0.0058}$  & ${0.1111}^{+0.0055}_{-0.0055}$ & ${0.0766}^{+0.0039}_{-0.0044}$ & ${0.1192}^{+0.0038}_{-0.0038}$ & $\mathbf{{0.0945}^{+0.0024}_{-0.0025}}$ \\
$\Omega_{\Lambda}$ & ${0.741}^{+0.095}_{-0.099}$  & $0.711^{+0.024}_{-0.025}$ & ${0.763}^{+0.010}_{-0.010}$ & ${0.694}^{+0.016}_{-0.016}$ & $\mathbf{{0.725}^{+0.007}_{-0.007}}$ \\
$\Omega_{m}$ & ${0.259}^{+0.099}_{-0.095}$  & ${0.289}^{+0.025}_{-0.024}$ & ${0.237}^{+0.010}_{-0.010}$ & ${0.306}^{+0.016}_{-0.016}$ & $\mathbf{{0.275}^{+0.007}_{-0.007}}$ \\
$n_{s}$ & ${0.964}^{+0.015}_{-0.015}$  & ${0.967}^{+0.014}_{-0.014}$ & ${0.973}^{+0.018}_{-0.016}$ & $\mathbf{{0.951}^{+0.011}_{-0.011}}$ & ${0.909}^{+0.010}_{-0.010}$ \\
$w$ & ${-1.12}^{+0.42}_{-0.43}$  & ${-0.92}^{+0.10}_{-0.10}$ & ${-0.65}^{+0.04}_{-0.03}$ & $\mathbf{{-1.04}^{+0.11}_{-0.11}}$ & ${-0.85}^{+0.02}_{-0.02}$ \\
$H_0(Km \cdot s^{-1} Mpc^{-1})$ & ${75}^{+15}_{-14}$  & ${68}^{+3.9}_{-2.7}$ & ${64}^{+0.2}_{-0.3}$ & $\mathbf{{68}^{+2.2}_{-2.1}}$ & ${64}^{+0.2}_{-0.2}$ \\
$age(Gyr)$ & ${13.75}^{+0.29}_{-0.27}$  & ${13.81}^{+0.11}_{-0.11}$ & ${14.34}^{+0.10}_{-0.10}$ & $\mathbf{{13.88}^{+0.1}_{-0.1}}$ & ${14.60}^{+0.09}_{-0.09}$ \\
$log[10^{10} A_s]$ & $-$  & ${3.078}^{+0.035}_{-0.035}$ & ${2.945}^{+0.040}_{-0.040}$ & $\mathbf{{3.086}^{+0.032}_{-0.033}}$ & ${2.942}^{+0.032}_{-0.032}$ \\[1ex]
\hline
\end{tabular}
\label{table:params}
\end{table}
\end{landscape}

\begin{figure}[!hbtp]
\centering
\includegraphics[angle=0, width=0.8 \textwidth]{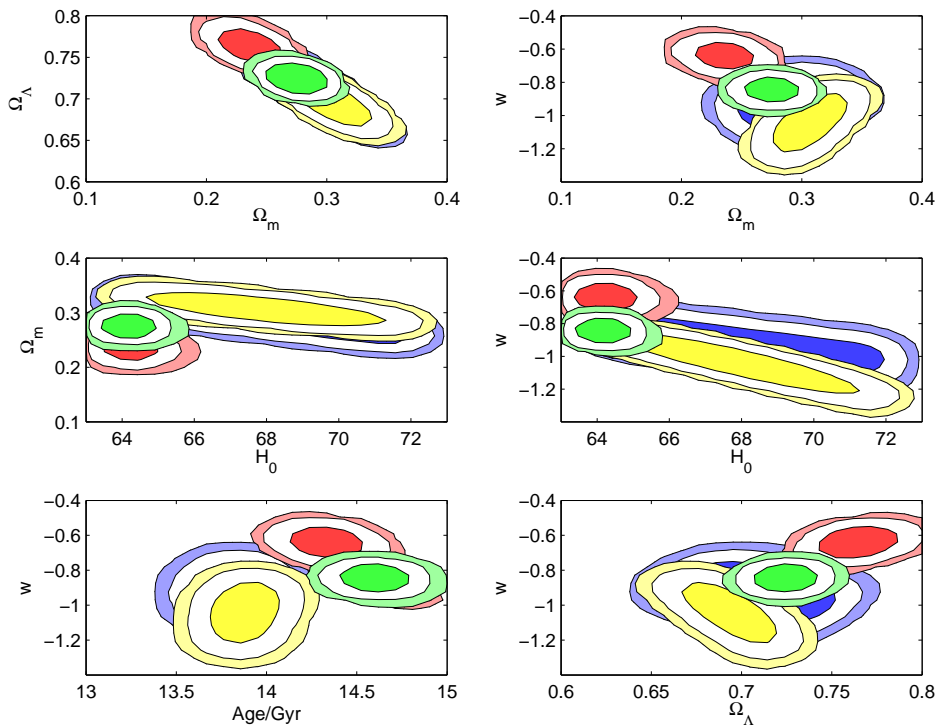}
\caption[Short]{\small Joint 2D marginalized distributions at 68\%, 95\%, 99\% CL showing the correlation between ($\Omega_{m}, \Omega_{\Lambda}$), ($\Omega_m , w$), ($H_0, \Omega_m$), ($H_0, w$), ($age, w$), ($\Omega_{\Lambda}, w$). The blue contour is for WMAP only data set, the red contour for WMAP+LBT, the yellow for WMAP7+Chandra and the green contour for WMAP7+LBT+Chandra data sets.}
\end{figure}
%.................................................................................

\section{Conclusions}
%..............................................................................
 The aim of our paper was to analyze how the LBT and Chandra data sets combined with WMAP7 measurements can improve the determination of the equation of state for the dark energy and also of the other parameters of the standard cosmological model.
We choose the LBT external data set because it contains information independent of each other and we can avoid biases present in the cosmic distance ladder method and the Chandra data set because it provides one of the best constrains on the matter density parameter $\Omega_m$.

 Our contribution consist in implementing LBT and Chandra new modules into the public Monte Carlo Markov Chain package. We run the modified CosmoMC package
for the following combinations of data sets: WMAP7, WMAP7+LBT, WMAP7+Chandra, WMAP7+LBT+Chandra. We found that, the physical baryonic density parameter, the scalar spectral index, the equation of state for the dark energy, the Hubble expansion rate, the age of the universe and the amplitude at the pivot scale are best constrain using WMAP7+Chandra data sets. All the other considered parameters, the physical dark matter, dark energy and matter density parameters and the plane $\Omega_{m} - \Omega_{\Lambda}$ are best constrain when we combine WMAP7+LBT+Chandra.

 Moreover, we conclude that the looking back time is a trustful measurement and in the future a larger LBT data set can further improve the determination of parameters we investigate.

 Acknowledgements. This work is supported by the ESA/PECS Contract "Scientific exploitation of Planck-LFI data" and by CNCSIS Contract 539/2009.

\end{document}